\documentclass[twocolumn,showpacs,preprintnumbers,amsmath,amssymb,prl]{revtex4}
\usepackage{graphicx}
\usepackage{bm}

\begin{document}

\title{Decoherence Rate of Semiconductor Charge Qubit\\ Coupled to Acoustic Phonon Reservoir}

\author{L. Fedichkin}
\author{A. Fedorov}
\affiliation{Center for Quantum Device Technology, Department of
Physics\\
and Department of Electrical and Computer Engineering,\\
Clarkson University, Potsdam, NY 13699--5720, USA}

\begin{abstract}
We analyze decoherence of an electron in a double-dot due to the
interaction with acoustic phonons. For large tunneling rates
between the quantum dots, the main contribution to decoherence
comes from the phonon emission relaxation processes, while for
small tunneling rates, the virtual-phonon, dephasing processes
dominate. Our results show that in common semiconductors, such as
Si and GaAs, the latter mechanism determines the upper limit for
the double-dot charge qubit performance measure.
\end{abstract}



\maketitle

Recently, there has been a lot of interest in implementation of
quantum logic gates by manipulating two-level electron systems in
semiconductor quantum dots (artificial atoms)~\cite{Barenco}.
Several designs for solid state quantum information processing
have been suggested~\cite{solid,Fedichkin,Dzurak}. Quantum-dot
architecture of a quantum computer is very attractive because it
is possibly scalable and the most compatible with the recent
microelectronics technology. However, it is a great challenge to
maintain a satisfactory level of coherence of an electron in
semiconductor to perform even elementary quantum
gates~\cite{Ekert}. Hopefully, coherence can be enhanced by
encoding of the logical qubit states into a subspace of the
electron states in a large quantum dot array (artificial
crystal)~\cite{Zanardi}. It is also noted that in a
gate-engineered structure of two coupled identical quantum dots
one can control decoherence rates by several orders of
magnitude~\cite{Fedichkin}. Recent advances in technology of
fabrication of double-dot~\cite{Fujisawa,Cain} and
double-donor~\cite{Dzurak} qubits have been reported. Coherent
oscillations in double-dot qubit are observed~\cite{Hayashi}. It
have been demonstrated that scattering by phonons can
significantly influence electron transport through double-dot
system~\cite{Brandes} and qubit dynamics during
measurement~\cite{Gurvitz}. In this work, we analyze decoherence
of an electron in a double-dot potential due to acoustic phonons
during one qubit gate cycle.

We consider a single electron in the double well potential shown
schematically in Fig.~1. Such a structure can be fabricated as two
 gate-engineered quantum dots~\cite{Fujisawa,Cain,Hayashi},
whose geometry is determined by the pattern of external metallic
gates and electric potential at them, or by the coupling the two
nearby phosphorus donors embedded in silicon~\cite{Dzurak}. The
resulting qubit is supposed to evolve in the basis spanned by the
states $|0\rangle$ and $|1\rangle$ which describe the electron
localized around the left and right minima of the potential,
respectively. We assume that the parameters of the double-dot
qubit structure are selected appropriately and the temperature is
low enough such that the effects of the electron transitions to
the higher energy levels can be neglected. Investigation of
decoherence due to acoustic phonons is the primary goal of our
work. Below, we will present the model and describe the two main
mechanisms of decoherence. We will introduce the appropriate
approximations schemes, quantify the overall error rate and
discuss the ways to minimize it.

\begin{figure}
\includegraphics[width=8.5cm, height=8cm]
{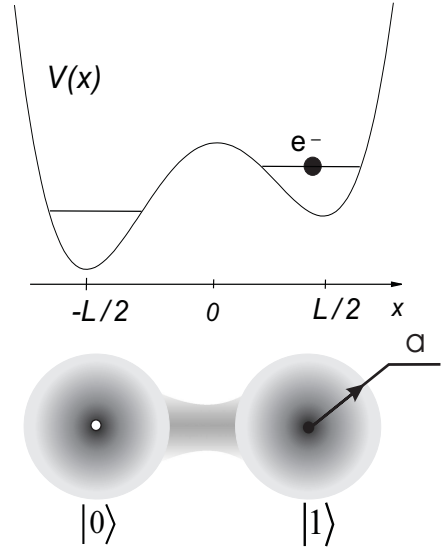} \caption{Sketch of the qubit: single electron within
double well potential.} \label{fig:1}
\end{figure}

The Hamiltonian of the electron and the phonon bath is
\begin{equation}
 H = H_e + H_p + H_{ep}.
\end{equation}
Here qubit term in Hamiltonian is
\begin{equation}\label{He}
 H_e = - \frac{1}{2}\varepsilon _{A}(t)\sigma _x- \frac{1}{2}\varepsilon _{P}(t)\sigma _z,
\end{equation}
where $ \sigma _x,\;\sigma _z$ are the Pauli matrices. The parameters
$\varepsilon _{A},\;\varepsilon _{P}$ are controlled by the
external metallic gates and can be used to perform on demand
single-qubit rotations. These parameters determine the splitting, $\varepsilon$, between the ground state
and the first excited state of the electron in the energy basis. This
splitting is given by $\varepsilon=\sqrt{\varepsilon
_{A}^2+\varepsilon _{P}^2}$.
The phonon
term in the Hamiltonian is
\begin{equation}
 H_p = \sum\limits_{\mathbf{q},\lambda } \hbar sq {\kern 1pt} b_{\mathbf{%
q},\lambda }^{\dagger}  b_{\mathbf{q},\lambda },
\end{equation}
where $ b^{\dagger}_{\mathbf{q},\lambda}$ and $
b_{\mathbf{q},\lambda}$ are the creation and annihilation operators of
the phonons with the wave vector $\mathbf q$ and polarization
$\lambda.$ For simplicity we consider isotropic acoustic phonons
with the linear dispersion law. The electron-phonon interaction
term is~\cite{Mahan}
\begin{equation}  \label{int}
 H_{ep}=\sum\limits_{\mathbf{q},\lambda }
 \sigma_z {\kern 1pt}\left( g_{\mathbf{q},\lambda }  b_{\mathbf{q},
\lambda}^{\dagger} +  g_{\mathbf{q},\lambda }^* b_{\mathbf{q},
\lambda}\right),
\end{equation}
where $g_{\mathbf{{q}, \lambda}}$ is the coupling constant, which
depends on the specific configuration of the system and the type
of the interaction. Both the distance $L$ between quantum dots centers and
their finite size $a$ will cut off the effect of the electron-phonon
interaction at the tails of the phonon spectrum.

One can show that for the interaction of an electron bound in a
gate-engineered Gaussian-shaped double-dot with deformation
phonons, the coupling constant is
\begin{equation}\label{7}
g_{\mathbf q}=i q \Xi\left( \frac \hbar{2\rho q s V}\right)
^{1/2}e^{-i\mathbf q\cdot \mathbf R-a^2q^2/4}\sin (\mathbf q\cdot
\mathbf L/2),
\end{equation}
where $\Xi$ is the deformation potential, $s$ is the speed of
sound, $\rho$ is the density of the crystal,  $V$ is normalizing
volume, and $\mathbf R$ is the coordinate of the middle point of
the double-dot. For crystal structures with inversion symmetry,
like Si, there is no additional interaction due to the
piezopotential. For crystals of the symmetry class $T_d$, like
GaAs, the piezoelectric phonon coupling is
\begin{eqnarray}  \label{p6}
g_{\mathbf q, \lambda}=&&-\left( \frac{\hbar}{2 \rho q s V
}\right)
^{1/2}Me^{-i\mathbf q\cdot \mathbf R-a^2q^2/4}\nonumber\\
&&\times(\xi _1e_2e_3+\xi _2e_1e_3+\xi _3e_1e_2)\sin (\mathbf
q\cdot \mathbf L/2),
\end{eqnarray}
where $e_i=q_i/q$, $\mathbf\xi$ is the polarization vector and $M$ is
the piezoconstant of the substrate.

For a double-donor system composed of two coupled hydrogen-like
dopant impurity states, e.g., for two phosphorus atoms embedded in
silicon, the following expression for the coupling constant with
the deformation phonons was obtained
\begin{equation}\label{hyg}
g_{\mathbf q}=i q\Xi \left( \frac \hbar{2\rho q s V}\right)
^{1/2}\frac{e^{-i\mathbf q\cdot \mathbf R}}{\left(1+a^2 q^2 /
4\right)^2}\sin (\mathbf q\cdot \mathbf L/2).
\end{equation}

The interaction term~(\ref{int}) leads to decoherence of the
qubit. The resulting loss of coherence is some functional of
$\varepsilon _{A}(t)$ and $\varepsilon _{P}(t)$.
Here we consider two representative cases of the single-qubit gate
functions and derive estimates for the error rate due to phonons.
First, we consider the
relaxation of an electron during the NOT gate ($\sigma_x$),
implemented by setting
$\varepsilon _{A}(t)=\varepsilon \equiv const$ and $\varepsilon
_{P}(t)=0$ in the Hamiltonian~(\ref{He}), for the time interval (cycle time of the quantum computer)
$\Delta t= \pi \hbar/\varepsilon $. Second, we consider the
decoherence of an electron during the $\pi$-phase-rotation gate ($\sigma_z$),
implemented by setting $\varepsilon _{A}(t)=0$ and
$\varepsilon _{P}(t)=\varepsilon \equiv const$ for the same time
interval $\Delta t=\pi \hbar/\varepsilon$.

To evaluate the relaxation of a double-dot qubit due to acoustic
phonons, we will follow~\cite{Fedichkin,Brandes,Barrett}. We
assume that the temperature is low compared to energy gaps of the
system. Therefore we consider the qubit at zero temperature. The
major parameter of dots influencing the interaction with phonons
is their size $a$. For gate-engineered quantum dots the actual
shape of wave function of confined electron can vary. We consider
Gaussian-shaped dots in which electron wave function is Gaussian
$\left(\Psi(\mathbf{r})\sim\exp{\left(-{r}^2/\left(2
a^2\right)\right)}\right)$. With these assumptions, the following
result for the relaxation rate due to the interaction with
deformation phonons can be obtained,
\begin{equation}\label{GDA}
 \Gamma_{DA}=\frac{\Xi^2k^3}{4\pi \rho s^2\hbar}\exp\left(-a^2k^2/2\right)\left(1-\displaystyle\frac{\sin\left(kL\right)}{kL}\right),
\end{equation}
where $k=\varepsilon/(s \hbar)$ is the wave-vector of the emitted
phonon. For the piezoelectric type of interaction, we get
\begin{widetext}
\begin{equation}\label{PDA}
 \Gamma_{PA}=\displaystyle\frac{M^2}{20\pi\rho s^2\hbar L^5 k^4}\exp{\left(-a^2
 k^2/2\right)}
\left(\left(kL\right)^5+5 kL\left( 2 \left(kL\right)^2 - 21\right) \cos\left(kL\right)+15\left(7 - 3\left(kL\right)^2\right)\sin\left(kL\right)\right).
\end{equation}
\end{widetext}

In double phosphorus dopant structures in silicon, the relaxation
rate due to the deformation phonons for the hydrogen-like impurity
states is
\begin{equation}\label{IDA}
\Gamma_{IDA}= \frac{\Xi^2}{4 \pi \rho
s^2\hbar}\frac{k^3}{\left(1+a^2 k^2 /
4\right)^4}\left(1-\frac{\sin (kL)}{kL}\right).
\end{equation}
If the wavelength of the phonon to be emitted is high enough compared
to the size of dots, $a$, and the distance between the dots, $L$, i.e.,
\begin{equation}\label{gg}
ak\ll1;\; Lk\ll1,
\end{equation}
which is often the case in present-day heterostructures, then the
following approximate expressions are valid
\begin{equation}\label{r01}
\Gamma_{DA}=\Gamma_{IDA}=\frac{\Xi^2L^2\varepsilon^5}{24\pi \rho
s^7 \hbar^6}
\end{equation}
and
\begin{equation}\label{r3}
\Gamma_{PA}=\displaystyle\frac{M^2L^2\varepsilon^3}{120\pi \rho
s^5\hbar^4}.
\end{equation}

Right after the implementation of the NOT gate the density matrix in
the energy basis $\left\{
\left|+\right\rangle,\left|-\right\rangle\right\}$, where
$\left|\pm\right\rangle=\left(\left|0\right\rangle \pm
\left|1\right\rangle\right)/\sqrt{2}$, will be~\cite{Blum}
\begin{equation}\label{m}
\left(%
\begin{array}{cc}
  1-\rho _{--} (0)e^{ - \Gamma \Delta t} & \rho _{+-} (0)e^{ - (\Gamma/2-i\varepsilon/\hbar )\Delta t} \\
  \rho _{-+} (0)e^{ - (\Gamma/2+i\varepsilon/\hbar) \Delta t} & \rho _{--} (0)e^{ - \Gamma \Delta t} \\
\end{array}%
\right),
\end{equation}
where $\rho _{\pm \pm} (0)$ are the elements of the electron density
matrix before the implementation of the NOT gate and the parameter
$\Gamma$ should be taken from Eqs.~(\ref{GDA}--\ref{IDA}),
respectively.

We now consider the implementation of the phase gate. In this
case, decoherence emerges as pure dephasing. There is no
relaxation because the interaction term~(\ref{int}) commutes with
the electron term in the Hamiltonian~(\ref{He}). The basis
$\left\{ \left|0\right\rangle,\left|1\right\rangle\right\}$
coincides with the energy basis of the electron. For evaluation of
the dephasing rate we used the general analytical expression for
the density operator of the electron in the boson bath given
in~\cite{vanKampen,Privman1},
\begin{equation}\label{mm}
\rho =\left(%
\begin{array}{cc}
  \rho _{00} (0)& \!\!\!\!\rho _{01} (0)e^{ - B^2(\Delta t)+i\varepsilon \Delta t/\hbar} \\
  \rho _{10} (0)e^{ - B^2(\Delta t)-i\varepsilon \Delta t/\hbar} & \!\!\!\!\rho _{11} (0) \\
\end{array}%
\right).
\end{equation}
Thus, the evolution of the system is determined by the spectral
function $B(t)$~\cite{Palma,vanKampen}, which in our case is
expressed as
\begin{equation}  \label{b2}
B^2(t)=\frac{8}{\hbar^{2}} {\sum\limits_{\mathbf q, \lambda} }
\frac{\left| g_{\mathbf q, \lambda}\right| ^2}{s^2 q^2} \sin ^2
\frac{s q t}2.
\end{equation}
By performing the summation in Eq.~(\ref{b2}), we get the spectral
functions determining the density matrices after the $\pi$-phase
rotation of both the qubits made of double-dots with deformation
and piezoelectric electron-phonon interaction, and of
double-impurity qubit states, respectively,
\begin{equation}\label{b2da}
B^2_{DA}=\frac{\Xi^2}{2\pi^2\rho s^3 a^2\hbar},
\end{equation}
\begin{equation}\label{b2pa}
B^2_{PA}=\frac{M^2\!L^2\!\left(1-\exp\left(-\displaystyle\frac{a^2\pi^2}{2L^2}\right)\!
+
\displaystyle\frac{3a^2}{L^2}E_1\left(\frac{a^2\pi^2}{2L^2}\right)\right)}{60\pi^2\rho
s^3a^2 \hbar},
\end{equation}
\begin{equation}\label{b2ida}
B^2_{IDA}=\frac{\Xi^2}{3\pi^2\rho s^3 a^2\hbar}.
\end{equation}
Here $E_1(z)=\int\limits_z^{\infty}t^{-1}e^{-t}dt$.
Expressions~(\ref{b2da}--\ref{b2ida}) were obtained by using an
additional observation that the duration of the qubit phase rotation is
large compared to the phonon transit time $\Delta t \gg a/s$. This
condition holds for the GaAs and Si structures considered.

To analyze the double-dot qubit architecture with respect to the
fault-tolerant quantum computing criteria~\cite{DiVincenzo}, one
should be able to estimate the error generated during the ''clock'' time
of the quantum computer $\Delta t$. To quantify the error due to
decoherence, we use the approach of Ref.~\cite{norm}. We consider the norm of the deviation operator,
$\sigma$
\begin{equation}\label{deviation}
  \sigma(t)  =   \rho(t)  -   \rho_{\rm ideal} (t),
\end{equation}\par\noindent
where the ''ideal'' evolution is defined as that at zero interaction with the environment,
\begin{equation}\label{f2}
\rho _{\rm ideal}(t)= e^{-i H_e t/\hbar}\rho(0)\, e^{i H_e t/\hbar}.
\end{equation}
\begin{figure}
\includegraphics[width=8.5cm, height=7.5cm]{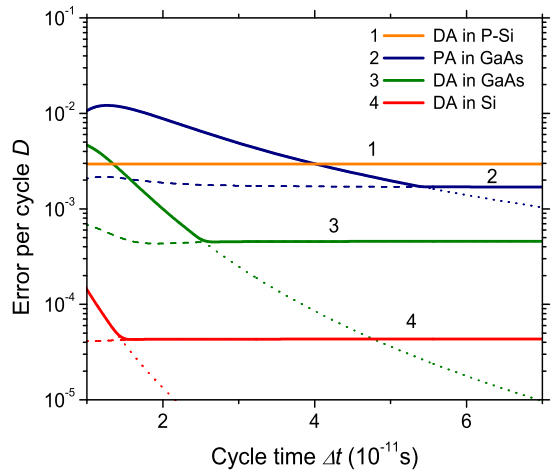}
\caption{Error rate estimate per cycle due to electron-phonon
interaction as a function of the cycle time $\Delta t$ ($\Delta
t=\pi \hbar/\varepsilon$). The distance between the dot centers
was $L=50~\rm{nm}$ for all the cases considered. For all the
gate-engineered quantum dots, the effective radius was
$a=25~\rm{nm}$. The parameters for the GaAs dots were
$\Xi=7~\rm{eV}$, $s=5.14~10^3~\rm{m/s}$, $\rho=5.31~\rm{g/cm^3}$,
$M=e e_{14}/(\epsilon_0 \kappa)$, where $e_{14}=0.16~\rm{C/m^2}$,
$\kappa=12.8$~\cite{Levinson}. As in Ref.~\cite{Barrett}, for
silicon the following parameters were used:  $a=3~\rm{nm}$ for
phosphorus impurity states, effective deformation potential
$\Xi=3.3~\rm{eV}$, $s=9.0~10^3~\rm{m/s}$, $\rho=2.33~\rm{g/cm^3}$.
The results are shown as follows: 1. The double-donor structure in
silicon; 2. Decoherence in GaAs due to piezointeraction; 3. The
contribution to decoherence due to deformation interaction in
GaAs; 4. The gate-engineered dots in Si. The dashed lines denote
$D_A$, the dotted lines denote $D_P$, the solid lines are $D$. The
relaxation rate for the double-donor structure in silicon is not
seen because it is very small in the given range of times.}
\label{fig:2}
\end{figure}

The error is characterized~\cite{norm} by the value
\begin{equation}\label{normD}
  D(t) = \sup_{\rho (0)}\bigg(\left\| \sigma (t,\rho (0))\right\|_{\lambda}
  \bigg),
\end{equation}\par\noindent
which is the maximal norm of the deviation operator over all the
possible initial density operators of the electron $\rho (0)$. For
fault-tolerant computation we need to satisfy the condition
$D(\Delta t)\leq O\left( 10^{-4}\right)$~\cite{DiVincenzo}. For
the density operators after the NOT and phase gates given by
Eqs.~(\ref{m}, \ref{mm}), the corresponding errors $D_A,D_P$ can
be expressed in a compact and elegant form as
\begin{equation}
 D_A(\Delta t)=1 - e^{ - \Gamma \Delta t},
\end{equation}
\begin{equation}
 D_P(\Delta t)  = \frac{1}{2}\left( 1 - e^{ - B^2
 (\Delta t)}\right).
\end{equation}
To evaluate the single-gate error rate we take the maximum of the
two gate errors considered which are typical single-qubit gates in
quantum algorithms. The error rate per each step can be estimated
as the larger of the above errors
\begin{equation}
D(\Delta t)=\max\left(D_A(\Delta t),D_P(\Delta t)\right).
\end{equation}\par\noindent
The obtained error rates for GaAs and Si quantum dots are shown in
Fig.~2. It should be noted that dephasing appears to be the
limiting factor of qubit fault tolerance. For a qubit made of
impurity states, the corresponding spectral function~(\ref{b2ida})
determining its dephasing rate is, in fact, material constant and
cannot be changed significantly. Since the estimated minimal error
rate for the Si double-impurity qubit is more than 1/400, the
construction of a practically useful fault-tolerant quantum
computer by using this design is questionable. Still, phonon
decoherence can be reduced by the change of phonon spectrum with
the help of phonon cavities~\cite{Fujisawa,Debald}.
Gate-engineered quantum dots show better coherence. Moreover,
their performance can be improved because their geometric
parameters are flexible.

In conclusion, we evaluated error rate in semiconductor charge
qubits due to interaction with acoustic phonons. Our results shows
that the expected error rate for double-phosphorus impurity states
in silicon is above the fault-tolerance threshold for quantum
computation. On the other hand, larger gate-engineered double
quantum dots both in Si and GaAs, with parameters close to those
in modern experiments~\cite{Hayashi,Cain,Dzurak}, can be
controlled more coherently. Realization of those qubits would be a
significant step toward the implementation of a full-scale solid
state quantum computer.

\begin{acknowledgments}
We gratefully acknowledge helpful discussions with Prof.
V.~Privman, D.~Mozyrsky and L.C.L.~Hollenberg. This research was
supported by the National Science Foundation, Grants DMR-0121146
and ECS-0102500, and by the National Security Agency and Advanced
Research and Development Activity under Army Research Office
Contract DAAD 19-02-1-0035.
\end{acknowledgments}


\begin{thebibliography}{20}

\bibitem{Barenco}
A.~Barenco, D.~Deutsch, A.~Ekert, and R.~Jozsa, Phys. Rev. Lett.
{\bf 74,} 4083 (1995).

\bibitem{solid}
D.~Loss and D.P.~DiVincenzo, Phys. Rev. A {\bf 57}, 120 (1998);
 A.~Imamoglu, D.D.~Awschalom, G.~Burkard, D.P.~DiVincenzo, D.~Loss, M.~Sherwin,
and A.~Small, Phys. Rev. Lett. {\bf 83}, 4204 (1999); V.~Privman,
I.D.~Vagner, and G.~Kventsel, Phys. Lett. A {\bf 239}, 141 (1998);
B.E.~Kane, Nature {\bf 393}, 133 (1998); R.~Vrijen,
E.~Yablonovitch, K.~Wang, H.W.~Jiang, A.~Balandin,
V.~Roychowdhury, T.~Mor and D.~DiVincenzo, Phys. Rev. A {\bf
62}, 012306 (2000).

\bibitem{Fedichkin}
L.~Fedichkin, M.~Yanchenko, and K.A.~Valiev, Nanotechnology {\bf
11}, 387 (2000); quant-ph/0006097.

\bibitem{Dzurak}
L.C.L.~Hollenberg, A.S.~Dzurak, C.~Wellard, A.R.~Hamilton, D.J.~
Reilly, G.J.~Milburn, and R.G.~Clark, cond-mat/0306235; A.S.~
Dzurak, L.C.L.~Hollenberg, D.N.~Jamieson, F.E.~Stanley, C.~Yang,
T.M.~Buhler, V.~Chan, D.J.~Reilly, C.~Wellard, A.R.~Hamilton,
C.I.~Pakes, A.G.~Ferguson, E.~Gauja, S.~Prawer, G.J.~Milburn,
 and R.G.~Clark, cond-mat/0306265.


\bibitem{Ekert}
A.~Ekert and R.~Jozsa, Rev. Mod. Phys. {\bf 68,} 733 (1996).

\bibitem{Zanardi}
P.~Zanardi, F.~Rossi, Phys. Rev. Lett. {\bf 81}, 4752 (1998).



\bibitem{Fujisawa}
T.~Fujisawa, T.H.~Oosterkamp, W.G.~van der Wiel, B.W.~Broer,
R.~Aguado, S.~Tarucha, and L.P.~Kouwenhoven, Science {\bf 282},
932 (1998).

\bibitem{Cain}
P.A.~Cain, H.~Ahmed, D.A.~Williams, J. Appl. Phys. {\bf 92}, 346
(2002), C.M.~Marcus et al., unpublished; C.G.~Smith, S.~Gardelis,
J.~Cooper, D.A.~Ritchie, E.H.~Linfield, Y.~Jin, and H.~Launois,
Physica E {\bf 12}, 830 (2002).

\bibitem{Hayashi}
T.~Hayashi, T.~Fujisawa, H.-D.~Cheong, Y.-H.~Jeong, Y.~Hirayama,
cond-mat/0308362.

\bibitem{Brandes}
T.~Brandes and B.~Kramer, Phys. Rev. Lett. \textbf{83},
3021(1999); T.~Brandes, F.~Renzoni, and R.H.~Blick, Phys. Rev. B
\textbf{64}, 035319 (2001); T.~Brandes and T.~Vorrath, Phys. Rev.
B \textbf{66}, 075341 (2002).


\bibitem{Gurvitz}
S.A.~Gurvitz, L.~Fedichkin, D.~Mozyrsky, and G.P.~Berman, Phys.
Rev. Lett. \textbf{91}, 066801 (2003).

\bibitem{Mahan} G.D.~Mahan, \emph{Many-Particle Physics\/}
(Kluwer Academic/Plenum Publishers, New York, 2000).

\bibitem{Barrett}
S.D.~Barrett and G.J.~Milburn, cond-mat/0302238.

\bibitem{Blum}
K.~Blum, \emph{Density Matrix Theory and Applications\/} (Plenum,
New York, 1996).

\bibitem{vanKampen} N.G.~van~Kampen, J. Stat. Phys. {\bf 78}, 299 (1995).

\bibitem{Privman1}
D.~Mozyrsky and V.~Privman, {J.\ Stat.\ Phys.} \textbf{91}, 787
(1998); quant-ph/9709020.

\bibitem{Palma}
G.M.~Palma, K.A.~Suominen, A.K.~Ekert,
{Proc. Roy. Soc. Lond. A} \textbf{452}, 567 (1996).

\bibitem{DiVincenzo}
D.P.~DiVincenzo, {Fort. Phys.} \textbf{48}, 771 (2000).

\bibitem{norm}
L.~Fedichkin, A.~Fedorov, and V.~Privman, Proc. SPIE
\textbf{5105}, 243 (2003); cond-mat/0303158.

\bibitem{Levinson}
V.F.~Gantmakher and Y.B.~Levinson, \emph{Carrier Scattering in
Metals and Semiconductors\/} (North-Holland, Amsterdam, 1987).

\bibitem{Debald}
S.~Debald, T.~Brandes, and B.~Kramer, Phys. Rev. B \textbf{66}, 
041301 (2002).

\end{thebibliography}
\end{document}